# DEPENDENCE OF E-CLOUD ON THE LONGITUDINAL BUNCH PROFILE: STUDIES IN THE PS & EXTENSION TO THE HL-LHC*

C. M. Bhat[#], Fermi National Accelerator Laboratory, Batavia, IL, 60510, USA
H. Damerau, S. Hancock, E. Mahner, F. Caspers, G. Iadarola, T. Argyropoulos
and F. Zimmermann, CERN, Geneva, Switzerland

*Abstract*

Recent studies have shown that the prospects for significantly increasing bunch intensities in the LHC for the luminosity upgrade (HL-LHC) may be severely limited by the available cryogenic cooling capacity and the electron-cloud (EC) driven beam instability. However, it is planned that during the HL-LHC era the bunch intensities in the LHC will go up by nearly a factor of two compared to the LHC-design values. This motivates the exploration of additional EC mitigation techniques that can be adopted in addition to those already in place. Preliminary simulations indicated that long "flat" bunches can be beneficial over Gaussian bunches to reduce the EC build up. Rigorous studies using realistic bunch profiles have never been done. Therefore, we have undertaken an in-depth investigation in the CERN 26 GeV PS to see if we can validate the previous findings and, in particular, if flattening the bunch can mitigate the EC. Here we present the results from dedicated EC measurements in the PS using a variety of bunch shapes and a comparison with simulations. Finally, we investigate if reshaping the bunch profiles using a $2^{nd}$ harmonic rf cavity can mitigate EC in the HL-LHC.

## I. INTRODUCTION

Issues related to the electron cloud in lepton and hadron circular accelerators have become a serious problem for future high-intensity upgrades. The primary source of the e-cloud in these accelerators are interactions of the circulating charged particle beam with residual gas (i.e., by gas ionization) and/or by interactions of synchrotron radiation emitted by the circulating beam with the walls of the accelerator beam pipe. The former mechanism is relevant in medium energy hadron accelerators like CERN PS, SPS, Fermilab Booster and Main Injector etc. On the other hand, the latter mechanism plays a major role in many lepton accelerators and high energy hadron accelerators like the LHC.

Since the first identification of an e-cloud induced beam instability in 1965 and its cure by implementing a transverse feedback system in a small proton storage ring of the INP Novosibirsk by Budker and co-workers [1], significant research has been carried out at various accelerator facilities around the world [2-5] to understand the EC dynamics and on the possible mitigation techniques. Addressing the EC related issues has become one of the important topics for designing new high intensity accelerators and for upgrading the beam intensities in the existing accelerators.

The Large Hadron Collider (LHC) [6] at CERN started physics operation in early 2010. Over the past two years tremendous progress has been made from the point of view of its performance. The design goal of the LHC luminosity was $1\times10^{34}$cm$^2$sec$^{-1}$ (with 25-ns bunch spacing) at a collision center of mass energy of 14 TeV. Currently, the LHC has reached more than 70% of its design peak luminosity at 57% of its design energy. For the High Luminosity LHC (HL-LHC) [7] two LHC bunch spacings – 25 ns and 50 ns – are under consideration. After the completion of the upgrade the peak luminosity (referred to as "peak virtual luminosity") is expected to be in excess of $20\times10^{34}$cm$^2$s$^{-1}$ and the bunch intensity to be increased by up to a factor of two.

At present, the LHC operates with a maximum of 1380 bunches with a bunch spacing of 50 nsec and intensities of about $1.5\times10^{11}$ppb. The experiments carried out in 2011-12 showed that EC-driven vacuum problems in the LHC [8] could be one of the major limiting factors for 25-ns bunch spacing. This is the case despite several EC mitigation measures which had been adopted in the LHC design, like saw-tooth pattern on the beam screen inside the cold dipole region, low secondary emission yield (SEY) NEG coatings on the inside surface of the warm beam pipes, etc. As a result, a major machine development campaign has been undertaken since 2011 to mitigate EC formation by beam scrubbing [9]. Consequently, significant improvement was seen [10] in the LHC performance. During the HL-LHC era the increased bunch intensity and the reduced bunch spacing will certainly aggravate EC related problems. Therefore, it is prudent to search for novel methods which could be complementary to beam scrubbing and can be used in combination with others to reduce EC formation.

Early simulation studies in the LHC indicated that there is an anti-correlation between increased bunch length and the electron cloud formation; very long bunches with rectangular profile can reduce EC considerably [11]. But such bunches are presently not being considered for any of the LHC upgrade scenarios. On the other hand, an in-depth analysis using realistic but nearly flat short bunches suitable for the LHC was never done. To shed light on this question, a dedicated EC experiment has been carried out in the CERN PS at ejection momentum of 26 GeV/c, where we investigated EC dependence on the shape of the bunch profiles. Fitting the EC simulations to the measurement data, we tried to study the correlation between bunch length and the EC evolution. Finally, we extrapolated our results and extended these studies to the HL-LHC scenarios.

High-intensity bunches in the HL-LHC also face an additional issue related to single and multi-bunch instabilities driven by the loss of the Landau damping [12]. Significant research has been carried out in the CERN SPS using its 4th harmonic rf system [13]. It has been concluded that operation of this higher harmonic rf system in the so-called bunch shortening mode renders the high-intensity beam more stable. Consequently, adding an 800 MHz Landau cavity is foreseen to stabilize high intensity beam in the LHC during the HL-LHC era [14]. The bunch-shortening mode implies a high peak line charge density of LHC bunches, which may not be favourable with regard to EC. Therefore, it is important to examine the implications of using a higher harmonic rf system in the HL-LHC from the EC point of view.

Since 2007, the CERN PS has been equipped with a purpose-designed, dedicated one-meter long EC monitor in the straight section (SS) 98 [15]. Figure 1 shows a schematic view of the detector. It has two identical 30 mm diameter button pickups on the upper part and a stripline-type electrode on the bottom of the vacuum chamber. The pickup detectors are shielded differently: BPU1 and BPU2 use 0.7 mm thick perforated stainless steel sheets (providing ≈ 10% transparency) and two grids (with about 37% and 23% transparency), respectively.

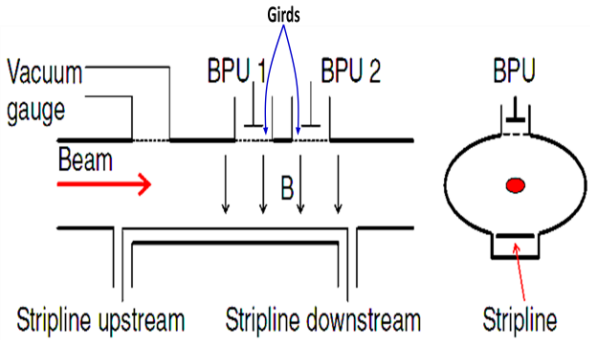

Figure 1: Schematic of the EC detector used in the PS straight section (SS) 98 (courtesy of E. Mahner [15]).

Clear EC signals and correlated vacuum degradation have been observed. The EC build up has been observed mainly for the last 36 ms before the beam ejection from the PS on the 25 nsec and 50 nsec bunch spacing LHC cycles [15]. Figure 2(a) shows the measured cumulative electrons from each pickup together with the vacuum pressure readings. Figure 2(b) shows typical PS mountain range [16] data during the last 140 ms on the same PS cycle. Figure 2(c) shows stages for rf turn-on times on the cycle (at flat-top) during the quadruple-splitting of the beam to finally produce a train of 72 bunches with 25 nsec bunch spacing. E. Mahner and his co-workers [15] have also deduced an approximate transfer function between the measured detector signals and the electron line density using system impedance, button transparencies etc,. They found that the relation between electron line density and button pickup voltage $U_{BPU1}$, is $\lambda/(e^-/m) = 2.3 \times 10^8 \, (U_{BPU1}/mV)$.

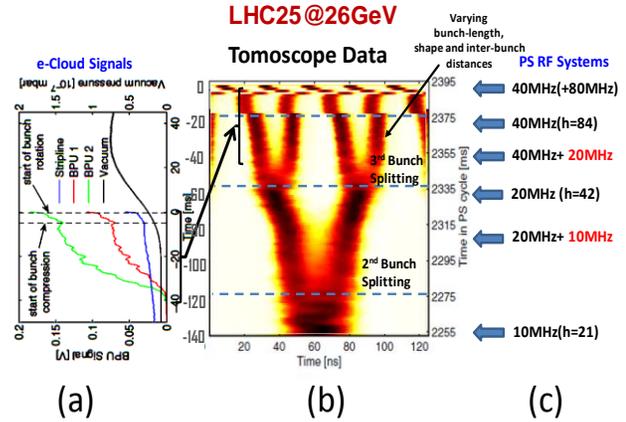

Figure 2: The region of interest from EC point of view in the PS beam on the LHC25 cycle[15] (for four bunches out of seventy two). (a) Measured EC signals from BPU1 (red curve), BPU2 (green curve) and stripline (blue curve) detectors along with vacuum (black curve) (b) mountain range data of the PS beam using tomoscope, and (c) used PS rf systems for beam quadruple splitting.

On the flat-top of the LHC25 cycle the bunch profile takes a variety of shapes and spans a range of bunch lengths. For example, at 40 ms before the ejection, the $4\sigma$ bunch length is about 15 ns as is shown in Fig. 3. During the final double splitting at about 60 ms before ejection (not shown in Fig. 3), dramatic bunch profile variation takes place in the double harmonic rf bucket made up of h=42 and h=84 rf systems. Eventually, an adiabatic bunch compression followed by a rapid bunch rotation (which is a quasi-nonadiabatic process) in a combined h=84 and h=168 rf bucket shortens the bunches to the final length of <4 ns at extraction. A very large growth in EC build up has been seen as the bunch rotation was taking place (see Fig. 2(a)). Fortunately, this spike in the EC density does not seem to have much detrimental effect on the PS beam because the latter is ejected exactly at this point on the cycle.

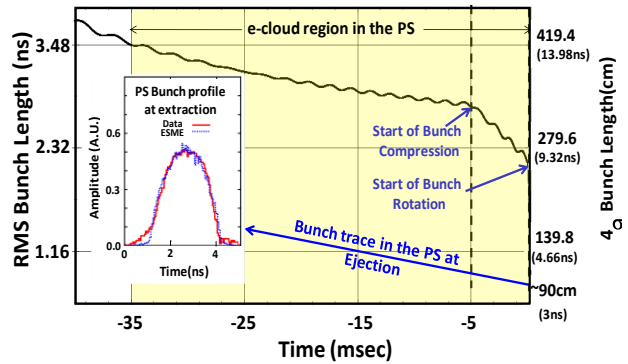

Figure 3: RMS bunch length variation during the last 40 ms on the PS-LHC25 beam cycle. The measured bunch profile just before ejection from the PS and its comparison with the predicted bunch profile using ESME is shown in the inset.

We realized that one can exploit the flexibilities of the PS in terms of rf system to investigate the EC effect for the bunch lengthening mode (BLM) and the bunch shortening mode (BSM) in a controlled environment with adiabatically changing bunch shapes, and then to conduct in-depth EC simulation studies to benchmark the available EC simulation codes against the measured data.

This paper is organized in the following way. We first give a brief review on the EC simulation codes used in the present analyses. In Sec. III, we discuss the dedicated EC experiment in the PS and the data analysis. Sec. IV describes the EC simulation effort for the HL-LHC operating scenario. In the final section we summarize our findings.

## II. E-CLOUD SIMULATIONS

The EC simulations have been carried out using ECLOUD [17] and a newly developed code PyECLOUD [18]. Both ECLOUD and PyECLOUD employ the same EC model, but the latter code uses faster algorithms and incorporates a few improvements. Both of these codes simulate EC cloud build up for the case when a train of bunches is injected into an empty accelerator section. The model adopted in both of these codes assumes that the total SEY, $\delta_{tot}$, is a sum of two quantities: i) a true SEY and ii) a component arising from elastic reflection. The sum is given by [3 (page 14), 4, 19],

$$\delta_{tot}(E_p,\theta) = \delta_{true}(E_p,\theta) + R_0\,\delta_{Elastic}(E_p) \quad (1)$$

where

$$\delta_{true}(E_p,\theta) = \delta_{Max}(\theta)\frac{sE_p/\varepsilon_{Max}(\theta)}{s-1+[E_p/\varepsilon_{Max}(\theta)]^s} \quad (2)$$

$$\delta_{Max}(\theta) = \delta_{Max}^{*}\exp[\frac{1}{2}(1-\cos(\theta))] \quad (3)$$

$$\varepsilon_{Max}(\theta) = \varepsilon_{Max}^{*}[1+0.7(1-\cos(\theta))] \quad (4)$$

$$\delta_{Elastic}(E_p) = \left(\frac{\sqrt{E_p}-\sqrt{E_p+E_0}}{\sqrt{E_p}+\sqrt{E_p+E_0}}\right)^2 \quad (5)$$

In the above equations the quantities $E_p$, $\delta_{true}$, $\delta_{Elastic}$, $\delta_{Max}$, $\varepsilon_{Max}$, $R_0$ and $\theta$, are the incident electron energy, the true secondary emission yield parameterized from the measurement data, the $E_p$-dependent elastic reflectivity (normalized so that $\delta_{Elastic} \to 1$ as $E_p \to 0$), the maximum of $\delta_{true}$, the incident electron energy at $\delta_{Max}$, the probability for elastic reflection in the limit of zero primary electron energy, and the angle of incidence of the primary electrons (with $\theta = 0$ taken to mean perpendicular impact), respectively, and, finally, with the two fitting parameters $E_0 = 150$ eV and $s \approx 1.35$ (a value of 1.35 has been determined for fully conditioned copper [19]). The quantity $R_0$ (in the range of 0 to 1) in this model accounts for a memory effect for the electrons inside the vacuum chamber even after the bunch train has passed by. In other words, the observed EC build up during the passage of a bunch train is enhanced by the passage of a preceding bunch train.

For most of the cycle the measured EC build up in the PS experiment [15] was in a steady-state condition (because, the rf manipulation was relatively slow compared to the EC growth and its decay per passage), except during the fast bunch rotation. In order to guarantee that a steady-state condition is reached in our simulated EC build up, it was necessary to carry out calculations for multiple passage of the PS bunch train taking into account the filling pattern, kicker gap and details of bunch profiles. In our simulations, we considered up to twenty passages for the same beam through the EC detector. (In Sec. III we will explain this aspect of the simulations in detail.)

Table 1: PS machine and EC parameters used in the ECLOUD and PyECLOUD simulations. Highlighted set is from best fitting to the measurement.

| Parameters | Values |
|---|---|
| Proton Momentum | 26 GeV/c |
| Number of Bunches/turn | 72 |
| Bunch Intensity | 1.35E11ppb |
| Bunch spacing | Varying (25-50nsec) |
| Bunch Length | Varying |
| Bunch Shape/Profiles | Varying shapes |
| Kicker Gap | 0.3 μs |
| Beam Pipe: H and V Aperture (half) | 7.3cm(H), 3.5cm(V) |
| Material of the Beam Pipe | Stainless Steel 316 LN |
| Beam Transvers Emit. $\varepsilon_x = \varepsilon_y$ | 2.1 μm |
| Lattice Function at the Detector $\beta x$ and $\beta y=$ | 22.14 m, 12.06 m |
| Ionization Crossection | 1 and 1.5 Mbarn |
| Gas Pressure | 10 nTorr |
| Maximum SEY yield $\delta_{Max}$ | 1.57 (Varied between 1.3-1.7) |
| R0: Probability for Elastic Reflection in the Limit of Zero Primary Energy of Electrons | 0.55 (Varied between 0.3-0.7) |
| Electron Energy at $\delta_{Max}$ (eV) | 287 (Varied bewteen 230-332) |

Table 1 lists the EC simulation parameters for the PS. Primary seed electrons are assumed to be produced by gas ionization. In our simulations we varied the gas ionization cross section by about 50% to investigate its effect on the saturation values of EC line-density. This study showed that the EC saturation value shows little dependence (<1%) on the ionization cross section for our beam and chamber parameters. The PS EC detector is located in an elliptical 316LN (low carbon with nitrogen) stainless steel chamber. Test-bench measurement data on the 316LN stainless steel [20] have been fitted to the non-linear curve described by Eq. (2) which gave $\delta_{Max}^{*} = 1.85$, $\varepsilon_{Max}^{*} = 282$ eV and $s=1.55$. These values are probably too pessimistic, because one may expect a significant reduction in the total SEY due to several years of beam scrubbing in the PS during its normal operation with LHC type beams. Therefore, we have carried out simulations searching for a somewhat reduced $\delta^*_{Max}$ in the range of 1.3 to 1.7 which best represents our data.

The EC simulations for the HL-LHC have been carried out only at the proton beam energy of 7 TeV and we assume that the primary seed electrons are exclusively due to the synchrotron-radiation induced photo-emission from the inner beam-pipe surface. In the model [19], about 80% of the photons produce photo-electrons when they first impact the beam pipe. All of these electrons lie in a narrow cone of $11.25^0$ and, in a strong dipole field, will never get much accelerated by the field of the proton beam. Consequently, they will not contribute to further EC build up. On the other hand, the photo-electrons produced by the remaining 20% of the photon flux are taken to be distributed azimuthally according to $\cos^2\psi$ and some of these contribute to the further EC build up in the LHC dipoles.

Table 2: HL-LHC machine parameters and EC parameters used in the ECLOUD and PyECLOUD simulations.

| Parameters | Values |
|---|---|
| Proton Energy | 7000 GeV |
| Number of Bunches/turn | 2808 @ 25nsec bunch spacing<br>1404 @ 50nsec bunch spacing |
| Bunch Intensity | 2.2E11ppb @ 25nsec bunch spacing 3.5E11ppb @ 50nsec bunch spacing |
| Bunch spacing | 25 and 50nsec |
| Bunch Length | Varying |
| Bunch Shape/Profiles | Varying shapes |
| Kicker Gap | 225nsec |
| Beam Pipe: H and V Aperture (half) | 2.2cm(H), 1.73cm(V) |
| Material of the Beam Pipe:<br>Warm sections<br>Cold sections | TiZrV Non-evaporable Getter (NEG) Coated<br>Cu-coated, Saw Tooth shapes |
| Beam Transvers Emit. $\varepsilon_x = \varepsilon_y$ | 2.5 $\mu$m for 25 nsec bunch spacing<br>3.0 $\mu$m for 50 nsec bunch spacing |
| Averge lattice function in simulations $\beta_x$ and $\beta_y=$ | 86.37 m, 92.04 m |
| Source of primary electrons & Reflectivity | 100% Photo emission<br>20% |
| Primary electron emission yield | 0.00087 |
| Reflected electron Distribution | $\cos^2\psi$ |
| Maximum SEY yield $\delta_{Max}$ | 1.3 to 1.7 |
| R0: Probability for Elastic Reflection in the Limit of Zero Primary Energy of Electrons | 0.2 t 0.7 |
| Electron Energy at $\delta_{Max}$ (eV) | 239.5 |

## III. PS E-CLOUD MEASUREMENTS

*Experiment*

The recent PS e-cloud measurements have been made using the PS EC detector and the PS beam cycle similar to the operational LHC25 cycle. Until 5 ms before beam extraction the rf manipulations have been kept unchanged. By this time, the final train of 72 bunches with 25 nsec bunch spacing was fully formed. The rf voltage of the 40 MHz rf system was programmed to be at 40 kV. The new rf manipulation sequences have been adopted as shown in Fig. 4(a). The 80 MHz rf system was turned on with the rf phase either at $0^0$ (in phase) or $180^0$ (counter phase). From here on, five different iso-adiabatic bunch manipulation schemes have been followed. 1) SH: voltage on the 40 MHz rf system has been increased linearly from 40 kV to 100 kV, keeping the 80 MHz rf system turned off. This left the bunches in a single harmonic rf bucket and the bunches were continuously being shortened for the next 5 ms (black curve in Fig. 4(a)). 2) BSM50: the 40 MHz and 80 MHz rf systems have been ramped up simultaneously in phase from 40 kV to 100 kV and 0 kV to 50 kV, respectively. Here the beam has been maximally squeezed giving rise to the shortest bunch and the final value of V2(80MHz)/ V1(40MHz)=0.5. 3) BSM25: similar to "2" but 80 MHz system ramped only up to 25 kV, 4) BLM25: similar to "3" but, rf systems in counter phase and 5) BLM50: similar to "2" but, rf systems in counter phase. This led to nearly "flat" bunches which results from V2(80MHz)/V1(40MHz)=-0.5.

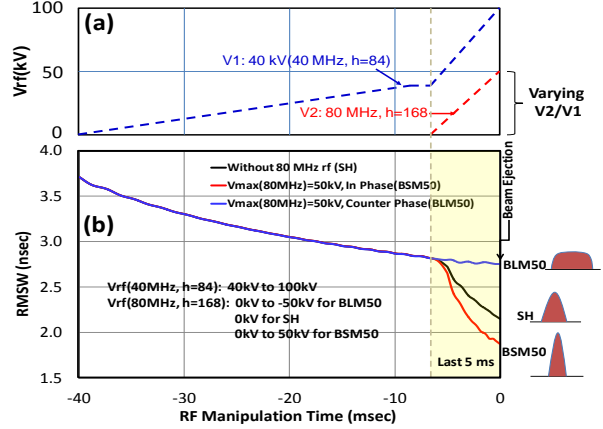

Figure 4: (a) PS rf manipulation and (b) ESME predicted bunch length variation during the last 40 ms before beam ejection. Until the last 5 ms the rf manipulations are identical to those of the operational cycle that produces bunches with 25 nsec spacing. During the last 5 ms, the 40 MHz and 80 MHz rf systems are ramped up simultaneously and linearly, to final values of 100 kV and 50 kV, respectively.

Figure 4(b) shows the simulated RMS bunch lengths in the PS for the entire rf cycles of interest using the longitudinal beam dynamics code ESME [21]. It is important to note that the rf voltage ratios V2(80MHz)/V1(40MHz) were varying from zero to a set final value of ± 0.50 during the rf manipulation period until the beam got ejected. Ideally, we wanted to hold the beam at the final values of the voltage ratios for an extended period. Operational constraints on the LHC25 cycle during the time of the experiment prevented this.

Figure 5 shows the measured bunch profiles using the PS tomoscope application for the region where EC build up is observed. The total PS beam intensities for the three cases shown here were 980x10¹⁰, 985x10¹⁰ and 973x10¹⁰ for BLM50, SH and BSM50, respectively. The average final bunch population was about 20% larger than that used in ref. 15. A total of 140 traces with delay of 480 PS revolution periods from trace to trace were recorded. The trace number and the corresponding time on the PS cycle relative to the beam ejection are listed in Table 3. Data show that the general features for all of the traces from 104 to 135 for the three different cases resemble each other except for a small difference arising from the beam intensity variation (<1%). Trace135 to Trace140 correspond to the last 5 ms and for these traces the bunch profiles of the three cases differ significantly. The measured RMS transverse emittance (inferred by wire scanners) was about 2.1 µm.

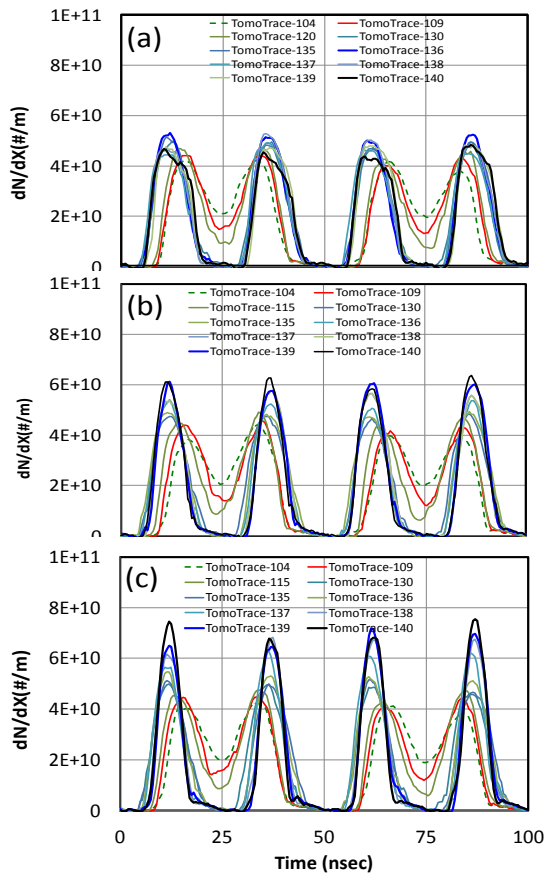

Figure 5: PS bunch profiles during the last 40 ms of the rf manipulations for a) BLM50, b) beam in h=84 rf buckets (SH) and c) BSM50 for four bunches out of 72. In these cases, the bunch rf manipulations differ only during the last 5 ms. The trace numbers in the figure indicate relative time in the PS cycle (see Table 3).

Figure 6(a) displays typical bunch profiles at beam ejection for all five cases studied here. The RMS bunch lengths in each case have also been listed for comparison. Figure 6(b) shows a typical PS bunch train of 72 bunches at ejection. The bunch to bunch intensity variation was <10%.

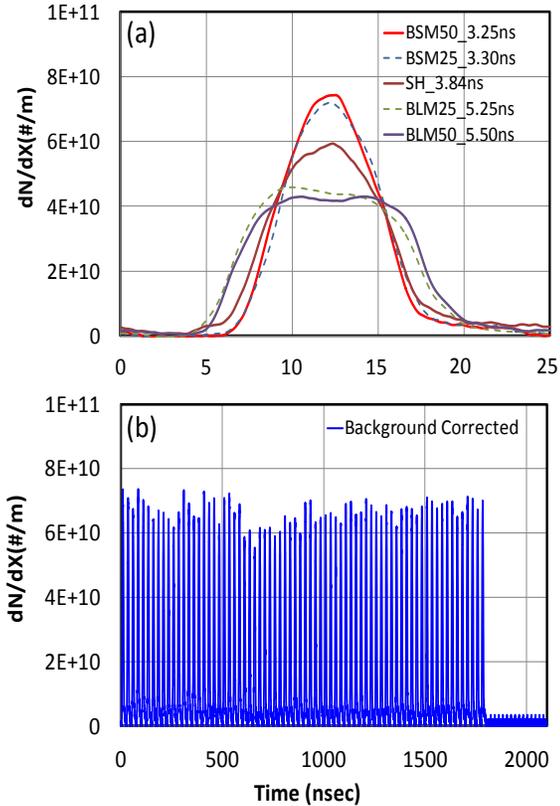

Figure 6: Typical PS bunch profile experimental data at ejection for a) all five cases studied here b) an illustration of entire train of 72 bunches (after background correction). The single bunch intensity was about $1.35\times10^{11}$ ppb in all the cases shown here.

Table 3: Trace number versus time relative to the beam ejection from the PS. These are referred to in Figure 5.

| Trace | Time Relative to PS Beam Ejection (ms) | Comments |
|---|---|---|
| Trace104 | -36.24 | Background |
| Trace109 | -31.21 | Start of EC |
| Trace115 | -25.17 | Growth pt.(Mid) |
| Trace120 | -20.13 | ≈Stable EC |
| Trace130 | -10.07 | Same as Above |
| Trace135 | -5.03 | 40MHz⊕80MHz |
| Trace136 | -4.03 | ,, |
| Trace137 | -3.02 | ,, |
| Trace138 | -2.01 | ,, |
| Trace139 | -1.01 | ,, |
| Trace140 | 0* | ,, |

*The fast bunch rotation was removed from the rf cycle on LHC25 during these experiments

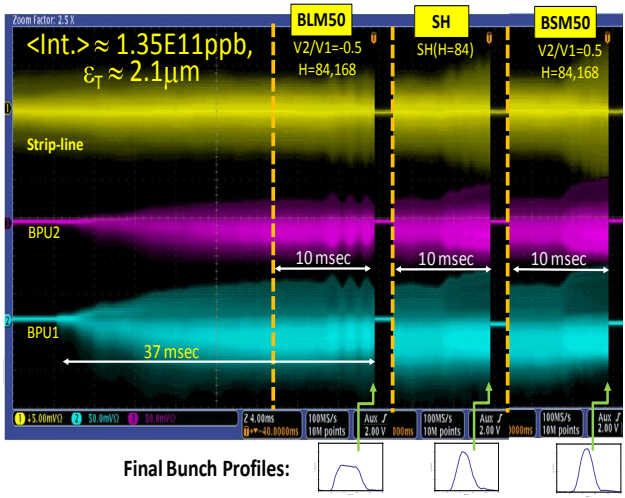

Figure 7: Signals from the EC monitor from three different detectors viz., strip-line, BPU1 and BPU2 for three rf manipulation scenarios. The bunch shapes at ejection are also shown.

Figure 7 presents typical EC monitor scope data over the last 37 ms on the PS cycle for BLM50 and data over the last 10 ms for the SH and BSM50 cases. Figure 8 shows the EC line density reconstructed from BPU1 for each of the PS turns with a bunch profile shown in Fig. 5. One can see a clear difference between the EC growth for BLM50 and the other two cases only during the last 5 ms. The data show that growth and saturation values strongly depend on the bunch profiles. However, independent of their peak electron-line density each one will decay in about 0.1 μsec after passage of the last bunch. Since the rf manipulations are sufficiently slow (i.e., the incremental change in bunch profile is almost negligible for a number of passages through the EC detector region as compared with EC growth and decay time, unlike in the case of fast bunch rotation mentioned in Sec. I), one can assume that the EC line density has reached a steady state in all cases shown in Fig. 8.

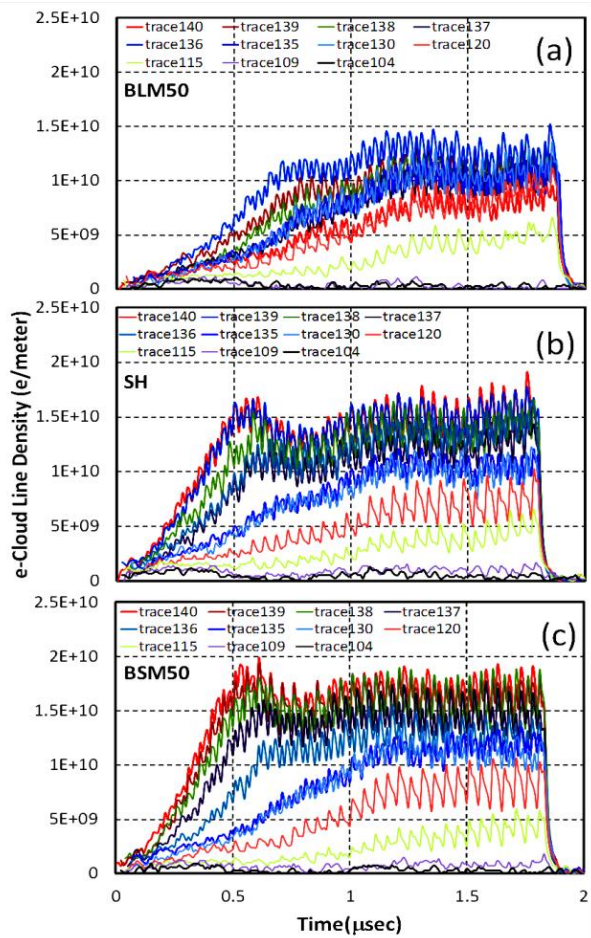

Figure 8: EC line-density measured at different time of the PS cycle during the last 40 ms before the beam ejection. The data shown from BPU1 are for a) BLM50, b) SH and c) BSM50. Notice that the EC behaviour was similar till trace130, but differs significantly from trace130 onward (also see Fig. 10 for Trace140).

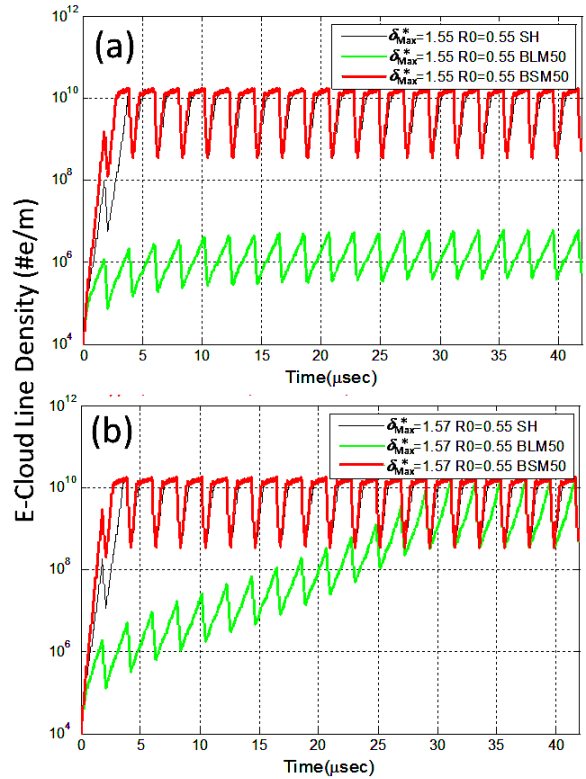

Figure 9: PS EC simulations using PyECLOUD with $\varepsilon^*_{Max}$ = 287 eV and a) $\delta^*_{Max}$ = 1.55, $R_0$=0.55; b) $\delta^*_{Max}$ = 1.57, $R_0$= 0.55 (optimized). Calculations are carried out for the drift section of the PS EC detector. These two cases are shown as examples to illustrate the combined sensitivity of EC growth on the SEY parameters and on the bunch shape.

## EC Simulations and Comparison with the Data

Initially, the simulation studies of the measured EC build up in the PS have been carried out using the code ECLOUD. The original version of the code could handle only standard Gaussian bunch profiles with a few non-standard shapes like flat, trapezium shapes etc. Also, there were issues related to adopting a non-standard filling pattern. The code has thus been modified to incorporate complex bunch profiles including a non-standard bunch filling pattern. In the meantime, PyECLOUD became available which could accommodate both standard as well as non-standard bunch profiles. All the simulation results presented here for the PS cases have been obtained with the PyECLOUD code.

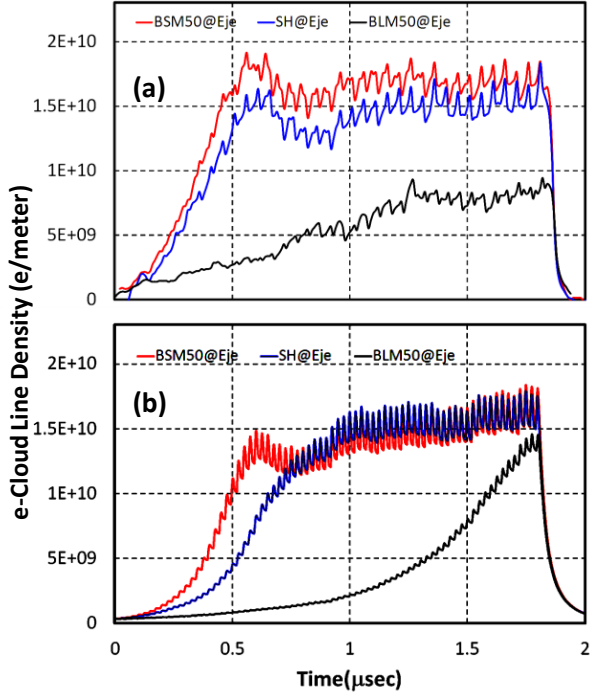

Figure 10: (a) Measured EC line-density in the PS at ejection and (b) the PyECLOUD simulations results corresponding to the cases shown in "a". The simulations have been carried out using high-lighted parameters in Table 1.

Starting from the measured values of $\delta^*_{Max}$ = 1.85 and $\varepsilon^*_{Max}$ = 282 eV for the 316LN stainless steel, we scanned the SEY parameter space (see Table 1). All of our simulations take the exact bunch profiles into account (shown in Fig. 5) with bunch to bunch intensity variation similar to that shown in Fig. 6(b) and the measured beam intensity in the PS. Figure 9 illustrates an example of such simulation results for two sets of SEY parameters and for three different beam profiles at ejection. The black, green and red curves are for the SH, BSM50 and BLM50 cases, respectively. For the cases shown in Fig. 9(b) the steady state was reached within about fifteen passages of the PS beam. In all of our simulations we allowed up to 20 passages. These simulations clearly show the sensitivity of the EC build up to the bunch profile and the SEY parameters. In the example of Fig. 9(a), we observe about four orders of magnitude change in EC line density for a 2% change in $\delta^*_{Max}$ between BLM50 and BSM50. This suggests that one could possibly use the bunch profile dependence of EC growth to estimate the SEY quite accurately.

Figure 10 displays a comparison between the measured and the simulated e-cloud line density (using $\varepsilon^*_{Max}$ = 287 eV, $\delta^*_{Max}$ = 1.57 and $R_0$ = 0.55) for the ejection traces. There is no normalization between the simulation results and the measurement data. We find quite a good agreement between the saturation values for the BSM50 and SH cases. Also, the overall trend is well reproduced. In the case of BLM50 the quality of the agreement is less satisfactory. Here the simulated EC line density grows rather slowly initially and then reaches a steady state maximum at a level about 30% higher than the measured value. However, as we will see next, even for this case the predicted cumulative number of electrons per turn lies within 30% of the measured value, of $3 \times 10^{12}$.

Next, simulations have been carried out using the same set of SEY parameters as mentioned above, to predict the complete EC build up through the experiment. Figure 11 presents the measured cumulative number of electrons per PS turn versus the relative time in the PS cycle. The 10% error assigned to the measured data points includes a systematic error and a background subtraction error. The three overlaid curves represent simulation results multiplied with a normalization factor of 0.85. The overall trend of the cumulative electrons is predicted quite well in all three cases. Simulations are found to reproduce even the observed oscillations during the last 5 ms in the case of BLM50. However, for SH and BSM50, the accumulated electrons on the last turn of the beam in the PS are underestimated by 25% and 50%, respectively, in the simulations.

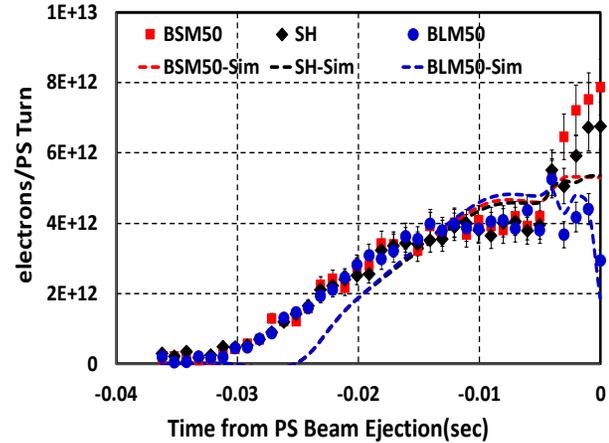

Figure 11: Overlay of the measured cumulative electrons/PS turn (red squares: BSM50; dark diamonds: SH; and blue circles: BLM50) and the predictions by PyECLOUD. The simulation data have been multiplied by a normalization factor 0.85 to better match the measurements (which could reflect a calibration error for the PS EC monitor).

From the PS study we clearly observe a dependence of EC growth on the bunch profile. We find the ratios BSM50/BLM50 ≈ 2.7±0.4 and SH/BLM50 ≈ 2.3 ±0.3 between the measured cumulative numbers of electrons at ejection. Certainly BLM results in considerably smaller EC build up than the other two cases. A comparison between measurements and simulations sets a tight range of values for the SEY parameters at the PS EC detector. For example, we found $\varepsilon^*_{Max}$ = 287 eV (± 3%), $\delta^*_{Max}$ = 1.57 (± 8%) and $R_0$ = 0.55 (± 3%). Also, we have been able to benchmark the EC simulation codes and the employed SEY model quite satisfactorily.

## IV. E-CLOUD IN THE HL-LHC

Over the last decade significant research has been carried out on the LHC EC issues [2-5, 8, 9, 19, 22 and 23]. Most of the past simulation studies assumed Gaussian bunch profiles and bunch intensities close to the LHC design values [6]. A lot of effort has been put into scanning the SEY parameter space. Ref. [19] presents EC-simulation results for the higher intensity operation of the LHC including some simulations for flat rectangular ~38 cm long (non-realistic to the LHC operating conditions) bunch profiles. Further, all of them have assumed about 25% and 50% larger transverse emittances for the 25-ns and 50-ns bunch filling patterns, respectively, than in the more recent HL-LHC specifications (Table 4). However, the EC is a very complex, non-linear multi-dimensional phenomenon. Further, the SEY parameters improve with machine operation. As a result of this, it is practically impossible to foresee every issue that one might encounter. In this section, we focus our study on realistic bunch profiles and better established SEY parameters.

Table 4: HL-LHC parameters of interest for EC issues [7]

| Parameter | Nominal | 25 ns Bunch spacing | 50 ns Bunch spacing |
|---|---|---|---|
| Beam Energy (TeV) | 7 | 7 | 7 |
| N (ppb)(xE11) | 1.15 | 2.20 | 3.50 |
| $n_b$(bunches per beam) | 2808 | 2808 | 1404 |
| Beam Current [A] | 0.58 | 1.12 | 0.89 |
| RMS bunch length (cm) | 7.55 | 7.55 | 7.55 |
| b-b Separation [s] | 9.5 | 12.5 | 11.4 |
| beta* at IP1&5 (m) | 0.55 | 0.15 | 0.15 |
| Normalized Emittance(μm) | 3.75 | 2.5 | 3 |
| X-Angle(mrad) | 285 (9.5s) | 590 | 590 |
| IBS rise time (z, x) [hr] | 57, 103 | 21, 15 | 16, 14 |
| Maximum Total b-b tune shift ($\Delta Q_{tot}$) | 0.011 | 0.015 | 0.019 |
| Peak Virtual luminosity [$10^{34}$ cm$^{-2}$s$^{-1}$] | 1 | 24 | 25 |
| Actual (leveled) pk luminosity [$10^{34}$ cm$^{-2}$s$^{-1}$] | 1 | 7.4 | 3.7 |
| Effective Beam lifetime[h] | 44.9 | 11.6 | 18.4 |
| Level time, run time | 0, 15.2 | 5.2, 8.9 | 11.4, |
| Beam Brightness [R.U.] | 1 | 2.9 | 3.8 |
| Pileup(@ Leveled Luminosity) | 19 | 140 | 140 |

Currently, the LHC is not instrumented with EC monitors as in the case of the PS and the SPS at CERN. All the information related to the EC in the LHC is deduced from the measured vacuum activities in various sectors of the ring and from the measured heat load in the cold arcs. Recently, a stringent range of SEY parameters has been deduced [10] by using the 2011-12 vacuum data in the uncoated warm regions of the LHC and comparing it with ECLOUD simulations, the parameters $\varepsilon^*_{Max}$ = 239.5 eV and $\delta^*_{Max}$ < 1.55 have been inferred. Here, we study the EC for the LHC using the HL-LHC beam parameters and the above values of SEY for a variety of possible realistic bunch profiles with the goal of investigating if a particular bunch profile is better than another from the point of view of EC mitigation.

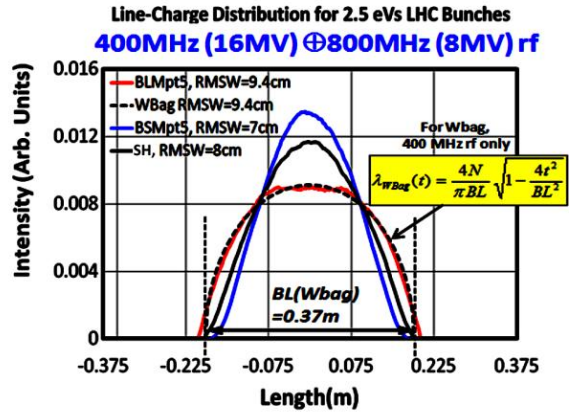

Figure 12: (ESME) Simulated HL-LHC beam bunch profiles in double harmonic rf buckets for BLM50 (BLMpt5), Waterbag, BSM50 (BSMpt5) and SH (in 400 MHz rf bucket).

Figure 12 shows ESME-simulated bunch profiles for the LHC. Guided by the measurements on the bunch profiles in the LHC at 4 TeV, we have used a Hofmann-Pedersen (elliptical) distribution for the beam in 400 MHz rf buckets at 7 TeV. An rf voltage of 16 MV is assumed. The profiles BSMpt5 and BLMpt5 have been generated by superposing the 2$^{nd}$ harmonic (800 MHz) rf wave on the fundamental rf wave of 400 MHz with V2/V1 = ±0.5, respectively. The dashed dark curve corresponds to the bunch profile from a "water-bag" model [24] (constant beam particle density distribution in the longitudinal phase space).

EC simulations have been carried out with ECLOUD as well as with the PyECLOUD using the parameters listed in Table 2 and 4. We have also extended some of the simulations to the intensity range of 1 to 4×10$^{11}$ppb. The current simulations use $\varepsilon^*_{Max}$ =239.5 eV, $\delta^*_{Max}$ in the range 1.3 to 1.7 and $R_0$ in the range of 0.2 to 0.7. We have considered a standard SPS batch of 288 bunches (similar to the one in the original LHC design) made of four batches from the PS (see for example Fig. 6(b)). The individual bunch profiles were similar to those shown in Fig. 12. For all values of SEY parameters used in our simulations, a clear signature of a steady state is seen by the end of the passage of the first PS batch as shown in Fig. 13. The SH profile has been used for both cases in this figure. Preliminary results from a similar EC simulation for the LHC with different bunch profiles generated using a double harmonic rf system have been reported earlier [26]. The electrons from EC, ultimately deposit their energy on the beam pipe. Heat load on the LHC cryo-system is due to the electron kinetic energy deposited on the beam pipe.

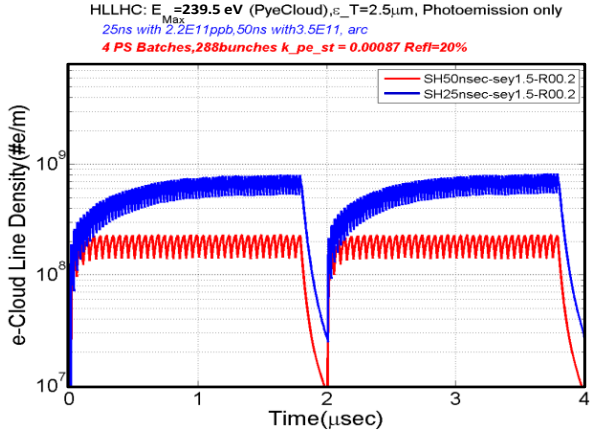

Figure 13: PyECLOUD simulations with $\varepsilon^*_{Max}$ =239.5 eV, $\delta^*_{Max}$ =1.5, $R_0$=0.2 for the HL-LHC beam parameters. Red and blue curves are for $3.5\times10^{11}$ppb with 50 nsec bunch spacing and $2.2\times10^{11}$ppb with 25 nsec bunch spacing, respectively. SH bunch profile is used in these simulations. For clarity, both of these curves are smoothened and results for only two PS batches are shown.

Cryogenic superconducting dipoles in the LHC occupy about 66% of the ring and carry the majority of the cryo-heat load. Therefore, we concentrate all of our simulations on the LHC dipoles (arcs). The calculated heat load for various bunch profiles and two sets of SEY are shown in Fig. 14(a) and the heat-load dependence on the bunch intensity is shown in Fig. 14(b). The contributions from quadrupoles and other cryo magnets to the total heat load are ignored here. The EC simulations for the arcs clearly show that the heat load has very little dependence on the bunch profiles. Therefore, BLM cannot be used as an EC mitigation technique in the LHC. The observed difference between PS and the LHC EC dependence on the bunch profiles may be primarily due to significantly shorter bunches in the LHC; the LHC bunches are about an order of magnitude smaller than those studied in the PS. For example, the shortest bunch in the PS (in our experiment) has a bunch length (4σ) of about 13 ns, while, for the LHC, the longest bunch length contemplated (4σ) is about 1.3 ns. Consequently, LHC bunches are too short to have any profile dependence on the EC growths. This aspect could be studied further.

Simulations show that even for the most pessimistic case of $\delta^*_{Max}$ = 1.7, $R_0$ = 0.7 (from Table 2) the average heat load is <0.5 W/m in the case of the 50 nsec bunch filling pattern. On the other hand, the calculated heat load for any of the 25-ns bunch filling patterns is more than the design heat-load handling capacity of the LHC cryo-system if $\delta^*_{Max}$ ≥1.5. Therefore, upgrades to the LHC cryo-system are inevitable for future operation with 25-ns bunch spacing at higher intensities unless the SEY is reduced significantly from the current values. Our simulations demonstrate that the LHC filling pattern with 50-ns bunch spacing has a clear advantage over the 25-ns bunch spacing even during the HL-LHC era.

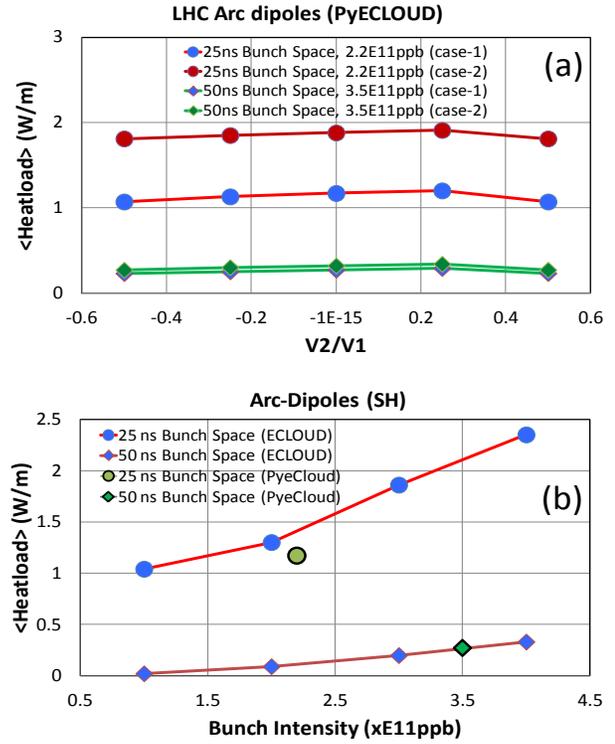

Figure 14: Calculated average heat load for the HL-LHC beam scenarios: a) bunch profile dependence (left-most points are for BML50 and rightmost points are for BSM50, the points at V2/V1 = 0 are for the SH). "case-1" implies $\delta^*_{Max}$ = 1.5, $R_0$ = 0.2. "case-2" implies $\delta^*_{Max}$ = 1.5, $R_0$ = 0.5. b) Bunch intensity dependence for "case-1" SEY parameters. ECLOUD simulations results are also shown for comparison.

The fact that the EC build up has little dependence on the bunch profiles in the LHC bodes well for the foreseen rf upgrades during the HL-LHC era. The high intensity beam can be made stable by use of a $2^{nd}$ harmonic Landau cavity if the bunches are in the BSM mode (or BLM mode for longitudinal emittance below some threshold [26]). With the current analysis, we show for the first time that the use of a Landau cavity in the LHC will have a negligible effect on the EC growth.

## V. SUMMARY

During the HL-LHC era the beam intensity in the LHC is expected to go up at least by a factor of two. This has direct implications on the EC growth and the issues related to the beam instability driven by the dynamics of the electron cloud. Therefore it is important to explore and develop techniques to mitigate EC growth. Fully developed techniques like NEG coatings on the inner surface of the beam pipe in warm sections and a saw tooth pattern on the beam screen inside the cold dipole region have been adopted in the LHC. Many new techniques are under consideration.

Early EC simulations have shown that the flat bunches have advantages over Gaussian bunches. In this regard, we conducted an EC experiment in the PS at its extraction energy where the EC is observed and the bunch profiles change significantly. Exploiting PS rf capabilities, a variety of possible bunch profiles, including nearly flat bunches, have been generated and the corresponding EC growth has been studied. Using the available EC codes at CERN, simulations have been carried out incorporating the measured PS bunch profiles. There was a good agreement between the EC measurements and the simulation results. These studies have enabled us to determine the SEY parameters for the EC monitor region of the PS quite accurately, as $\varepsilon^*_{Max}$ = 287 eV ($\pm$ 3%), $\delta^*_{Max}$ = 1.57 ($\pm$ 8%) and $R_0$ = 0.55 ($\pm$ 3%). We also find that the nearly flat (BLM50) bunches produce about a factor 2.7$\pm$0.4 lower number of electrons than Gaussian bunches.

We have then extended similar studies to the HL-LHC beam conditions through simulations, where the bunch lengths were nearly ten (3.25 nsec(in the PS during current experiment)/0.31 nsec(LHC)) times smaller than that in the PS at extraction. We found that in the LHC the EC growth is almost independent of bunch profiles. Consequently, the foreseen installation of a second harmonic Landau cavity, that would change bunch profiles to BSM and make the beam longitudinally more stable, will not pose any additional EC related problems in the LHC.

## ACKNOWLEDGMENT

The authors would like to thank the CERN operation team for their help while conducting the beam experiments in the PS accelerator. One of the authors (CMB) is specially indebted to O. Brüning, G. Arduini, E. Shaposhnikova, R. Garoby, G. Rumolo and S. Gilardoni for their hospitality at CERN and many useful discussions. Also, special thanks are due to Humberto M. Cuna for his help in early stages of simulation studies and M. Goodman for his comments on this paper. This work is supported by Fermi Research Alliance, LLC under Contract No. DE-AC02-07CH11359 with the U.S. Department of Energy, US LHC Accelerator Research Program (LARP) and Coordinated Accelerator Research in Europe – High Intensity, High Energy, Hadron Beam (CARE-HHH).